\documentclass[iop,apj]{emulateapj}

\usepackage{amsmath}
\usepackage{verbatim}
\usepackage{wasysym}
\usepackage{multirow}
\usepackage{soul} %% text highlighting, \hl command...
\usepackage[dvips]{color}

\newcommand\T{\rule{0pt}{2.6ex}}       % Top strut
\newcommand\B{\rule[-1.2ex]{0pt}{0pt}} % Bottom strut

\begin{document}

\title{Multi-Layer Hydrostatic Equilibrium of Planets and Synchronous Moons: Theory~and Application to Ceres and to Solar System Moons}

\author{Pasquale Tricarico}
\affil{Planetary Science Institute, Tucson, AZ 85719, USA}

\begin{abstract}
The hydrostatic equilibrium of multi-layer bodies lacks a satisfactory theoretical treatment 
despite its wide range of applicability.
Here we show that by using the exact analytical potential of homogeneous ellipsoids
we can obtain recursive analytical solutions and an exact numerical method for 
the hydrostatic equilibrium shape problem of multi-layer planets and synchronous moons.
The recursive solutions rely on the series expansion of the potential in terms of the polar and equatorial shape eccentricities,
while the numerical method uses the exact potential expression.
These solutions can be used to infer the interior structure of planets and synchronous moons
from the observed shape, rotation, and gravity.
When applied to dwarf planet Ceres, we show that it is most likely a differentiated body 
with an icy crust of equatorial thickness 30--90~km 
and a rocky core of density 2.4--3.1~g/cm$^3$.
For synchronous moons, we show that the 
$J_2/C_{22} \simeq 10/3$
and the $(b-c)/(a-c) \simeq 1/4$ ratios have significant corrections of order $\Omega^2/(\pi G \rho)$,
with important implications on how their gravitational coefficients are determined from flyby radio science data
and on how we assess their hydrostatic equilibrium state.
\end{abstract}

\keywords{planets and satellites: interiors --- planets and satellites: individual (Ceres)}

\maketitle

\section{Introduction}

Understanding how gravity, pressure, and rotation contribute to the shape of a 
homogeneous fluid body has been a remarkable achievement,
with many contributions over several centuries \citep{1969efe..book.....C}.
The homogeneous fluid body theory can be qualitatively applied
to large planets and moons \citep{1976eioh.book.....J},
but their differentiated interior structure leads to significant deviations
between theory and observations.
The differentiation of large bodies is a natural consequence 
of radiogenic heating \citep{1955PNAS...41..127U,1995Metic..30..365M,1998Icar..134..187G},
causing partial or total melting and segregation
of the heavier components towards the center of the body shortly after formation.
Tidal dissipation can also represent an important heating source,
as in the case of Io \citep{1979Sci...203..892P}.
This motivates us to investigate the hydrostatic equilibrium of multi-layer bodies.

The linear superposition of rotational and tidal deformations, 
which is a good approximation only for very slow rotators,
led \cite{1979Icar...37..575D} 
to compute the equilibrium figure of 2-layer planets and synchronous moons,
including the deformation of the interior layer.
A numerical 2-layer model by \cite{1993Icar..105..326T} was used in 
\cite{2005Natur.437..224T} to determine the interior structure of Ceres,
and we perform a similar analysis in \S\ref{SEC:CERES}.
In \cite{2010JGRE..11512003K} the 2-layer problem for planets is approached using 
spheroidal coordinates, and this leads to implicit integral equations
which are solved numerically,
with realistic examples presented in \cite{2011PEPI..187..364S}.
A recursive numerical form of the solution of the gravitational field of $N$-layer spheroids
is presented in \cite{2013ApJ...768...43H}.

In this manuscript 
we show how the analytic expressions for the potential of a homogeneous ellipsoid
can be used to obtain recursive high-order analytical solutions to the $N$-layer hydrostatic equilibrium problem,
with closed-form 2nd order equations.
Numerical methods can also be obtained with an accuracy which depends primarily on the precision of the floating point operations.
Applications to Ceres and to synchronous moons of the giant planets are presented.

\section{Methods}

In an incompressible $N$-layer fluid body in hydrostatic equilibrium,
pressure $p$, potential $U_\text{tot}$ and density $\rho$ 
satisfy the gradient equation
\begin{align}
\nabla p = \rho \nabla U_\text{tot}
\label{eq:hydro}
\end{align}
where the potential $U_\text{tot}$ 
has the form:
\begin{align}
U_\text{tot} & = U_\text{rot} + U_\text{tid} + \sum_{i=1}^{N} \frac{\rho_i-\rho_{i-1}}{\rho_i} U_i \label{eq:Utot} \\
U_\text{rot} & = \frac{x^2+y^2}{2} \Omega^2 \\
U_\text{tid} & = \frac{2x^2-y^2-z^2}{2} \Omega^2
\end{align}
where $U_\text{rot}$ is the rotational potential
in the body-fixed co-rotating frame,
$\Omega = 2\pi/T$ is the angular velocity,
and $T$ is the rotation period.
For synchronous moons,
$U_\text{tid}$ is the leading term of the tidal potential
\citep{1999ssd..book.....M},
where we have included only the static component.
Rotation is about the $z$ axis, and for synchronous moons the perturbing planet is along the $x$ axis.
We neglect the time-dependent component of the tidal potential as the moon is assumed to 
have negligible obliquity and 
to be on an orbit with negligible eccentricity.
$U_i$ is the potential of the $i$-th layer, which is either $U_\text{TE}$, in the generic case of a triaxial ellipsoid,
or $U_\text{OS}$ for oblate spheroids,
with detailed expressions provided in the Appendix.
Each layer $i$ is treated as a triaxial ellipsoid with semi-axes $a_i \geq b_i \geq c_i$,
polar eccentricity $e_{pi}^2 = 1 - (c_i/a_i)^2$, equatorial eccentricity $e_{qi}^2 = 1 - (b_i/a_i)^2$,
and density $\rho_i$.
In the sum of Eq.~\eqref{eq:Utot}, $i=1$ corresponds to the outer layer, 
and the density $\rho_i$ of each layer increases with $i$, while the layer size decreases
so that the layer $i+1$ is fully contained in the layer $i$:
$a_{i+1} \leq a_i$, $b_{i+1} \leq b_i$, $c_{i+1} \leq c_i$.
Outside the body, the density is $\rho_0=0$.
Finally, the contribution to $U_\text{tot}$ by each layer is proportional to its relative density increase,
as expressed in the factor $(\rho_i-\rho_{i-1})/\rho_i$.

To solve the multi-layer hydrostatic equilibrium problem of Eqs.~\eqref{eq:hydro} and \eqref{eq:Utot}
and determine the semi-axes of all the layers,
given their density and volume, 
it is sufficient to require that surfaces of constant density are equipotential.
In this problem, equipotential surfaces are approximated by coaxial ellipsoids.
This approximation is an excellent one:
in the interior of an isolated layer the equipotential surfaces are exact ellipsoids,
and as we show in Eq.~\eqref{eq:deviation_from_ellipsoid} in Appendix \S\ref{SEC:DEVIATION}
in the exterior of an isolated layer the deviation of an equipotential surface from an ellipsoid
is very small and decreases very rapidly with distance,
while it is identically zero along the principal axes.
The rotational nature of the problem imposes to all layers to be coaxial.
To verify that surfaces are equipotential it is then sufficient to compare the value of $U_\text{tot}$ at the three extremes along the principal axes.

\subsection{Analytical Solutions \label{SEC:ANALYTICAL}}

Analytical solutions of the multi-layer hydrostatic equilibrium problem
can be obtained in the form of recursive relations.
The principal idea of recursive equations originated while  
inverting the Maclaurin relation, which relates polar eccentricity $e_p$ 
and angular velocity $\Omega$ of a body with density $\rho$ in hydrostatic equilibrium
\citep{1969efe..book.....C}.
As we show in Appendix~\S\ref{SEC:INVML},
in the limit of small $e_p$ we can expand the Maclaurin relation in power series,
and then obtain a recursive relation which effectively inverts the Maclaurin
relation, providing $e_p$ as a function of $\Omega$ and $\rho$.
In order to apply this approach to the multi-layer hydrostatic equilibrium problem,
we use $U_\text{tot}$ from Eq.~\eqref{eq:Utot} 
where the potential $U_i$ of each layer is given by the power series expansions
in Eq.~\eqref{eq:U_TE_in_exp} and \eqref{eq:U_TE_out_exp}.
We then impose the condition for equipotential surfaces 
$U_\text{tot}(a_i,0,0) = U_\text{tot}(0,b_i,0) = U_\text{tot}(0,0,c_i)$
and isolate the polar and equatorial eccentricities $e_{pi}$ and $e_{qi}$
in recursive relations.

\subsubsection{Multi-Layer Planets \label{SEC:MULTI_PLANETS}}

For a 2-layer planet,
including terms up to the 2nd order in eccentricity in the gravitational potential
from Eq.~\eqref{eq:U_TE_in_exp} and \eqref{eq:U_TE_out_exp},
we have:
\begin{align}
e_{p1}^2 & \longleftarrow \frac{15\Lambda^2 + 12  \mu_2^5 \sigma_2 e_{p2}^2}{8 + 20 \mu_2^3 \sigma_2} + \cdots \label{eq:ep1sq_OS_N2_e2} \\
e_{p2}^2 & \longleftarrow \frac{15\Lambda^2 + 12 e_{p1}^2}{20 + 8\sigma_2} + \cdots \label{eq:ep2sq_OS_N2_e2}
\end{align}
where 
$\mu_i=a_i/a_1$, 
$\sigma_i=(\rho_i-\rho_{i-1})/\rho_1$, 
$\Lambda^2=\Omega^2/(\pi G \rho_1)$,
$G$ is the gravitational constant,
and we have $\mu_1=1$, $\sigma_1=1$.
The left hand eccentricities in Eq.~\eqref{eq:ep1sq_OS_N2_e2} and \eqref{eq:ep2sq_OS_N2_e2} are iteratively updated 
with the value of the expressions on the right,
with initial values $e_{pi} = 0$.
Higher order recursive relations are provided in Appendix~\S\ref{SEC:HIGH_RECURSIVE}.
For a 3-layer planet, the 2nd order recursive equations are:
\begin{align}
e_{p1}^2 & \longleftarrow \frac{15 \Lambda^2 + 12 \mu_2^5 \sigma_2 e_{p2}^2 + 12 \mu_3^5 \sigma_3 e_{p3}^2}{8 + 20 \mu_2^3 \sigma_2 + 20 \mu_3^3 \sigma_3} + \cdots \\
e_{p2}^2 & \longleftarrow \frac{15 \Lambda^2 + 12 e_{p1}^2 + 12 (\mu_3/\mu_2)^5 \sigma_3 e_{p3}^2}{20 + 8 \sigma_2 + 20 (\mu_3/\mu_2)^3 \sigma_3} + \cdots \\
e_{p3}^2 & \longleftarrow \frac{15 \Lambda^2 + 12 e_{p1}^2 + 12 \sigma_2 e_{p2}^2}{20+20\sigma_2+8\sigma_3} + \cdots
\end{align}
For a $N$-layer planet the general recursive relation for the $i$-th layer is then:
\begin{align}
e_{pi}^2 & \longleftarrow \frac{\displaystyle 15 \Lambda^2 + 12 \sum_{k=1}^{i-1} \sigma_k e_{pk}^2 + 12 \sum_{k=i+1}^{N} (\mu_k/\mu_i)^5 \sigma_k e_{pk}^2}{\displaystyle 20 \sum_{k=1}^{i-1} \sigma_k + 8 \sigma_i + 20 \sum_{k=i+1}^{N} (\mu_k/\mu_i)^3 \sigma_k} + \cdots
\label{eq:rot_N_epsq}
\end{align}
where the sums are to account for the effects of outer ($k<i$) and inner ($k>i$) layers relative to the $i$-th layer considered.
Once the shape of each layer has been determined,
we can obtain the inertia moments from Appendix~\S\ref{SEC:INERTIA}
and the expansion of the gravitational potential from Appendix~\S\ref{SEC:SH}.

The 2nd order recursive equations above converge to closed form equations,
which can be obtained by solving for the $N$ unknowns $e_{pi}$
the system of $N$ linearly independent equations $U_\text{tot}(a_i,0,0) = U_\text{tot}(0,0,c_i)$.
The 2-layer 2nd order solution in Eq.~\eqref{eq:ep1sq_OS_N2_e2} and \eqref{eq:ep2sq_OS_N2_e2}
converges to the closed form
\begin{align}
e_{p1}^2 & = 
\frac{15 \Lambda^2}{8}  
\frac{(1 + (2/5) \sigma_2 + (3/5) \mu_2^5 \sigma_2 )}{F_p} \label{eq:ep1_sq_planet} \\
e_{p2}^2 & = 
\frac{15 \Lambda^2}{8}  
\frac{(1 + \mu_2^3 \sigma_2)}{F_p} \label{eq:ep2_sq_planet}
\end{align}
where 
\begin{align*}
F_p = 1 + (2/5) \sigma_2 + (5/2) \mu_2^3 \sigma_2 + \mu_2^3 \sigma_2^2  - (9/10) \mu_2^5 \sigma_2
\end{align*}
For the ratio of the eccentricities, we have:
\begin{align}
\frac{e_{p2}^2}{e_{p1}^2} 
& = \frac{1 + \mu_2^3 \sigma_2}{1 + (2/5) \sigma_2 + (3/5) \mu_2^5 \sigma_2}
\end{align}
which is non-zero even in the limit of a small core:
\begin{align}
\frac{e_{p2}^2}{e_{p1}^2} & \xrightarrow{\mu_2 \to 0}   \frac{5 \rho_1}{3\rho_1+2\rho_2} \leq 1
\end{align}

The expressions for the $J_2 = -C_{20}$ gravity coefficient and the principal moment of inertia $C$ are then:
\begin{align}
J_2 & = \frac{3 \Lambda^2}{8} \frac{(1+(2/5)\sigma_2+(8/5)\mu_2^5\sigma_2+\mu_2^8\sigma_2^2)}{(1+\mu_2^3\sigma_2) F_p} \\
\begin{split}
\frac{C}{M a_1^2} & = \frac{2}{5} \frac{(1+\mu_2^5\sigma_2)}{(1+\mu_2^3\sigma_2)} \\
& - \frac{3 \Lambda^2}{20} \frac{\mu_2^3 \sigma_2^2 (1-\mu_2^2-(5/2)\mu_2^3+4\mu_2^5-(3/2)\mu_2^7)}{(1+\mu_2^3\sigma_2)^2 F_p} 
\end{split} 
\end{align}
The upper bounds can be obtained in the limit of an homogeneous body ($\sigma_2 \to 0$), and are
$ e_{p1} \leq (15/8) \Lambda^2 $,
$ J_2 \leq (3/8) \Lambda^2$ 
and 
$ C/(M a_1^2) \leq 2/5$.

\subsubsection{Multi-Layer Synchronous Moons \label{SEC:MULTI_MOONS}}

The presence of a tidal potential causes non-zero equatorial eccentricities $e_{qi}$,
and for a 2-layer synchronous moon,
including terms up to the 2nd order in eccentricity,
we have:
\begin{align}
e_{p1}^2 & \longleftarrow \frac{60\Lambda^2 + 12 \mu_2^5 \sigma_2 e_{p2}^2}{8 + 20 \mu_2^3 \sigma_2} + \cdots \\
e_{p2}^2 & \longleftarrow \frac{60\Lambda^2 + 12 e_{p1}^2}{20 + 8\sigma_2} + \cdots\\
e_{q1}^2 & \longleftarrow \frac{45\Lambda^2 + 12 \mu_2^5 \sigma_2 e_{q2}^2}{8 + 20 \mu_2^3 \sigma_2} + \cdots\\
e_{q2}^2 & \longleftarrow \frac{45\Lambda^2 + 12 e_{q1}^2}{20 + 8\sigma_2} + \cdots
\end{align}

Higher order recursive relations are provided in Appendix~\S\ref{SEC:HIGH_RECURSIVE}.
For a 3-layer synchronous moon, the 2nd order recursive equations are:
\begin{align}
e_{p1}^2 & \longleftarrow \frac{60\Lambda^2 + 12 \mu_2^5 \sigma_2 e_{p2}^2 + 12 \mu_3^5 \sigma_3 e_{p3}^2}{8 + 20 \mu_2^3 \sigma_2 + 20 \mu_3^3 \sigma_3} + \cdots \\
e_{p2}^2 & \longleftarrow \frac{60\Lambda^2 + 12 e_{p1}^2 + 12 (\mu_3/\mu_2)^5 \sigma_3 e_{p3}^2}{20 + 8 \sigma_2 + 20 (\mu_3/\mu_2)^3 \sigma_3} + \cdots \\
e_{p3}^2 & \longleftarrow \frac{60\Lambda^2 + 12 e_{p1}^2 + 12 \sigma_2 e_{p2}^2}{20 + 20 \sigma_2 + 8 \sigma_3} + \cdots \\
e_{q1}^2 & \longleftarrow \frac{45\Lambda^2 + 12 \mu_2^5 \sigma_2 e_{q2}^2 + 12 \mu_3^5 \sigma_3 e_{q3}^2}{8 + 20 \mu_2^3 \sigma_2 + 20 \mu_3^3 \sigma_3} + \cdots \\
e_{q2}^2 & \longleftarrow \frac{45\Lambda^2 + 12 e_{q1}^2 + 12 (\mu_3/\mu_2)^5 \sigma_3 e_{q3}^2}{20 + 8 \sigma_2 + 20 (\mu_3/\mu_2)^3 \sigma_3} + \cdots \\
e_{q3}^2 & \longleftarrow \frac{45\Lambda^2 + 12 e_{q1}^2 + 12 \sigma_2 e_{q2}^2}{20 + 20 \sigma_2 + 8 \sigma_3} + \cdots
\end{align}

For a $N$-layer synchronous moon the general recursive relation for the $i$-th layer is then:
\begin{align}
e_{pi}^2 & \longleftarrow \frac{\displaystyle 60 \Lambda^2 + 12 \sum_{k=1}^{i-1} \sigma_k e_{pk}^2 + 12 \sum_{k=i+1}^{N} (\mu_k/\mu_i)^5 \sigma_k e_{pk}^2}{\displaystyle 20 \sum_{k=1}^{i-1} \sigma_k + 8 \sigma_i + 20 \sum_{k=i+1}^{N} (\mu_k/\mu_i)^3 \sigma_k} + \cdots \label{eq:tidal_N_epsq} \\
e_{qi}^2 & \longleftarrow \frac{\displaystyle 45 \Lambda^2 + 12 \sum_{k=1}^{i-1} \sigma_k e_{qk}^2 + 12 \sum_{k=i+1}^{N} (\mu_k/\mu_i)^5 \sigma_k e_{qk}^2}{\displaystyle 20 \sum_{k=1}^{i-1} \sigma_k + 8 \sigma_i + 20 \sum_{k=i+1}^{N} (\mu_k/\mu_i)^3 \sigma_k} + \cdots \label{eq:tidal_N_eqsq}
\end{align}

Similarly to \S\ref{SEC:MULTI_PLANETS} 
the 2nd order recursive equations admit closed form solutions,
and the 2-layer 2nd order solution in Eq.~\eqref{eq:ep1sq_OS_N2_e2} and \eqref{eq:ep2sq_OS_N2_e2}
converges to
\begin{align}
e_{p1}^2 & = \frac{15 \Lambda^2}{2} \frac{(1 + (2/5) \sigma_2 + (3/5) \mu_2^5 \sigma_2)}{F_p} \label{eq:sat_ep1sq} \\
e_{p2}^2 & = \frac{15 \Lambda^2}{2} \frac{(1 + \mu_2^3 \sigma_2)}{F_p} \label{eq:sat_ep2sq} \\
e_{q1}^2 & = \frac{45 \Lambda^2}{8} \frac{(1 + (2/5) \sigma_2 + (3/5) \mu_2^5 \sigma_2)}{F_p} \label{eq:sat_eq1sq} \\
e_{q2}^2 & = \frac{45 \Lambda^2}{8} \frac{(1 + \mu_2^3 \sigma_2)}{F_p} \label{eq:sat_eq2sq}
\end{align}
Note how the polar eccentricities in Eqs.~\eqref{eq:sat_ep1sq}--\eqref{eq:sat_ep2sq}
and the equatorial eccentricities in Eqs.~\eqref{eq:sat_eq1sq}--\eqref{eq:sat_eq2sq}
are respectively a factor 4 and a factor 3 larger than 
the rotation-only polar eccentricities in Eqs.~\eqref{eq:ep1_sq_planet}--\eqref{eq:ep2_sq_planet}
for the corresponding layer,
a result originally attributed to Clairaut.

The expressions for the $J_2$ and $C_{22}$ gravity coefficients and the principal moment of inertia $C$ are:
\begin{align}
J_2 & = \frac{15 \Lambda^2}{16} \frac{(1+(2/5)\sigma_2+(8/5)\mu_2^5\sigma_2+\mu_2^8\sigma_2^2)}{(1+\mu_2^3\sigma_2) F_p} \\
C_{22} & = \frac{9 \Lambda^2}{32} \frac{(1+(2/5)\sigma_2+(8/5)\mu_2^5\sigma_2+\mu_2^8\sigma_2^2)}{(1+\mu_2^3\sigma_2) F_p} \\
\frac{C}{M a_1^2} & = \frac{2}{5} \frac{(1+\mu_2^5\sigma_2)}{(1+\mu_2^3\sigma_2)} - \frac{9 \Lambda^2}{8} \frac{G_C}{(1+\mu_2^3\sigma_2)^2 F_p} 
\end{align}
where
\begin{align*}
G_C & = 1 + (2/5)\sigma_2 + \mu_2^3\sigma_2 + (8/5)\mu_2^5\sigma_2 + (4/3)\mu_2^3\sigma_2^2 \\
& - (14/15)\mu_2^5\sigma_2^2 - (7/3)\mu_2^6\sigma_2^2 + (19/3)\mu_2^8\sigma_2^2 \\ 
& - (7/5)\mu_2^{10}\sigma_2^2 + \mu_2^{11}\sigma_2^3
\end{align*}
The upper bounds given by $\sigma_2 \to 0$ are
$e_{p1} \leq (15/2) \Lambda^2$,
$e_{q1} \leq (45/8) \Lambda^2$,
$ J_2 \leq (15/16) \Lambda^2$,
and
$C_{22} \leq (9/32) \Lambda^2$.
For $C/(M a_1^2)$ we have that it is smaller than the homogeneous moon value of
$(2/5) - (9/8) \Lambda^2$
for most values of $\mu_2$ except when $\mu_2 \simeq 1$, in which case 
$(2/5) - (9/8) \Lambda^2 \leq C/(M a_1^2) \leq 2/5$.

\subsection{Numerical Solutions \label{SEC:NUMERICAL}}

For fast-rotating multi-layer planets and synchronous moons
the polar and equatorial eccentricities $e_{pi}$ and $e_{qi}$ obtained with
analytical methods in \S\ref{SEC:ANALYTICAL} tend to converge very slowly to the exact solution,
even when using high-order recursive equations,
see Table~\ref{tab:two_layer_comparison} in \S\ref{SEC:COMPARISON}.
Additionally, the Maclaurin equation includes 
disk-like solutions with $e_{pi} \simeq 1$ \citep{1969efe..book.....C}
which are beyond the range of applicability of
methods based on power series expansion of the potential.
To overcome these limitations, we have included a numerical method
capable of finding all the admissible solutions,
including the high $e_{pi}$ ones and the Jacobi branch for triaxial ellipsoid solutions of planets.

In the numerical method we use $U_\text{tot}$ from Eq.~\eqref{eq:Utot} 
where the potential $U_i$ of each layer is given by 
the exact analytic expressions in Eq.~\eqref{eq:U_TE} and \eqref{eq:U_OS}.
We then impose the condition for equipotential surfaces 
$U_\text{tot}(a_i,0,0) = U_\text{tot}(0,b_i,0) = U_\text{tot}(0,0,c_i)$
by performing numerical minimization of $\Delta^2$:
\begin{align}
\begin{split}
\Delta^2 & = \sum_{k=1}^{N} \left[ U_\text{tot}(0,b_i,0) - U_\text{tot}(a_i,0,0) \right]^2 \\
         & + \sum_{k=1}^{N} \left[ U_\text{tot}(0,0,c_i) - U_\text{tot}(a_i,0,0) \right]^2
\end{split}
\end{align}
and solve for the polar and equatorial eccentricities $e_{pi}$ and $e_{qi}$.
The accuracy of the solution depends on the precision of the floating point operations,
including the effect of roundoff errors. 
Satisfactory solutions will have $\Delta^2$ compatible with zero within the numerical precision.
The initial values of $e_{pi}$ and $e_{qi}$ in general can be chosen at random between 0 and 1,
or can be seeded using the 2nd order recursive relations in Eq.~\eqref{eq:rot_N_epsq} for planets
or in Eq.~\eqref{eq:tidal_N_epsq} and \eqref{eq:tidal_N_eqsq} for synchronous moons when applicable.

\subsection{Comparison with Previous Works \label{SEC:COMPARISON}}

Our analytical and numerical solutions are now compared to previous results in the literature.
In \cite{2010JGRE..11512003K} the 2-layer problem is studied in spheroidal coordinates,
and numerical solutions are obtained along with some test cases. In particular,
we use their Figure~4 test case to perform a direct comparison with our solutions.
Note that in \cite{2010JGRE..11512003K} the convention for layer numbering is inverted with respect to ours.
In the 2-layer model, the polar eccentricity of each layer is determined while
varying the relative volume fraction of the core $Q_V=\mu_2^3 (1-e_{p2}^2)^{1/2} (1-e_{p1}^2)^{-1/2}$ from 0 to 1,
at a fixed density ratio $\rho_2/\rho_1=2$, or $\sigma_2=1$.
The rotation is fixed at $\Lambda^2=\Omega^2/(\pi G \rho_1)=2 \varepsilon_2$ with $\varepsilon_2=0.05$.
\begin{figure}
\begin{center}
\includegraphics*[width=\columnwidth]{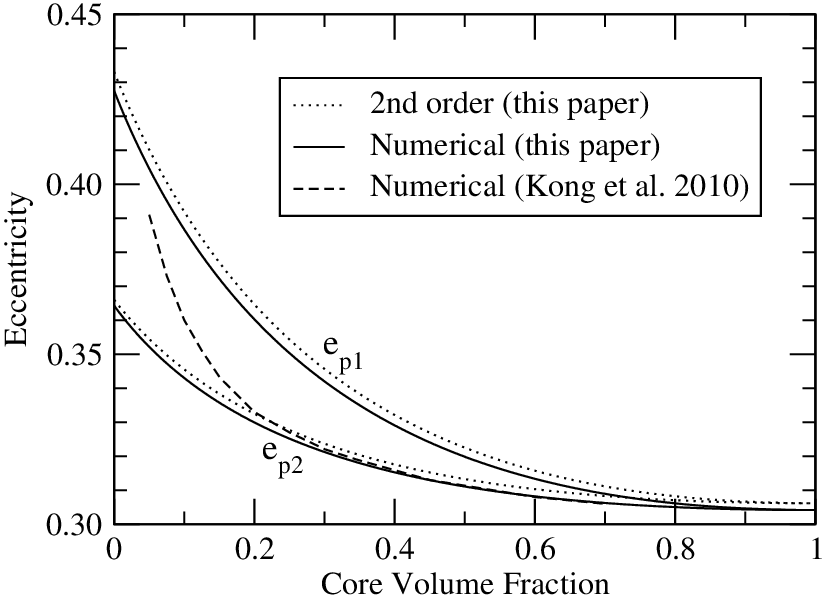}
\end{center}
\caption{Comparison of 2-layer solutions with Fig.~4 in \cite{2010JGRE..11512003K}.
The numerical solutions are indistinguishable for $e_{p1}$,
while they are significantly different for $e_{p2}$ for a small core.}
\label{fig:comparison_Kong2010}
\end{figure}
In Figure~\ref{fig:comparison_Kong2010} we show our results along with that of \cite{2010JGRE..11512003K}.
If we take our numerical solutions as reference, our 2nd order analytical solution is within approximately 1\%,
and the 4th order solution is within approximately 0.01\% and is not displayed in Figure~\ref{fig:comparison_Kong2010} 
because indistinguishable from the numerical one.
Our numerical values for the eccentricity of the outer layer $e_{p1}$ 
are in very good agreement with \cite{2010JGRE..11512003K}
over the whole range of core volume fraction, 
including the two limiting values of 0.4275 (no core)
and 0.3042 (all core).
However, for the eccentricity of the inner layer $e_{p2}$ we have good agreement only in the limit of a large core,
while for a small core the solution of \cite{2010JGRE..11512003K} 
seems to significantly over-estimate the core eccentricity, 
reaching a relative difference of over 10\% at $Q_V=0.05$.

Additional test cases are presented in \cite{2011PEPI..187..364S},
including their Figure~2 with the eccentricity of equipotential surfaces plotted against the equatorial radius.
\begin{figure}
\begin{center}
\includegraphics*[width=\columnwidth]{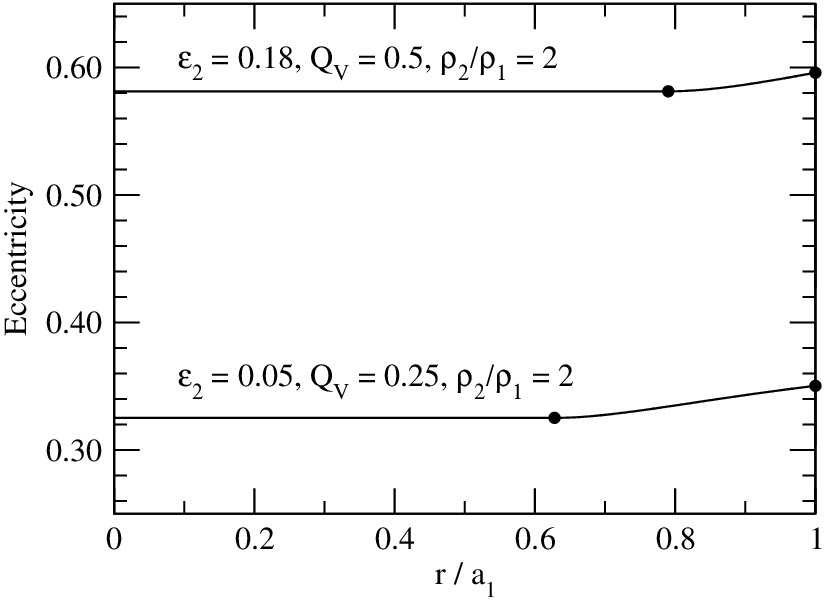}
\end{center}
\caption{Our 2-layer solutions for the test case described in Fig.~2 of \cite{2011PEPI..187..364S}.
The dots represent the equatorial radius and eccentricity of each layer.}
\label{fig:comparison_Schubert2011}
\end{figure}
Our results for this case are displayed in Figure~\ref{fig:comparison_Schubert2011},
where we have that the eccentricity of equipotential surfaces is constant within the core 
and then grows monotonically,
in contrast with what found by \cite{2011PEPI..187..364S} who
describe significant oscillations in proximity of the interface
and a general decrease when moving towards the center of the body.
In support of our results we have the following argument,
which applies in general to any number of layers and to both the rotational and tidal cases:
from Eq.~\eqref{eq:Utot} we have that 
the total potential in the innermost layer is given by
the exact expressions for $U_\text{TE,in}$ or $U_\text{OS,in}$ (in the Appendix) for each layer,
plus the rotational and tidal components,
and as such the total potential it is quadratic in the cartesian coordinates with constant factors,
because $\kappa=0$, so it generates equipotential surfaces with constant $e_p$ and $e_q$.

\begin{table}
\begin{center}
\begin{tabular}{l|lll}
Test Case & $Q_V$ & $\rho_1/\rho_2$ & $\varepsilon_2$ \T \B \\
\hline
Mars     & 0.125     & 0.486     & 0.00347   \T \\
Neptune  & 0.091125  & 0.157334  & 0.0254179 \\
Uranus 2 & 0.0563272 & 0.0791231 & 0.0318902 \B \\
\end{tabular}
\end{center}
\caption{Test cases and corresponding numerical constants used as input.}
\label{tab:two_layer_test_cases}
\end{table}

\begin{table*}
\begin{center}
\begin{tabular}{ll|lll}
Test Case & Method & $e_{p1}$ & $e_{p2}$ & $J_2 \times 10^6$ \T \B \\
\hline
Mars & \cite{2011PEPI..187..364S}     & \underline{0.100} 30              & \underline{0.088 8}59             & \underline{182}3.1           \T \\
Mars & \cite{1978ppi..book.....Z}     & \underline{0.100 29}5             & \underline{0.088 87}4 7           & \underline{182}3.18          \\
Mars & \cite{2013ApJ...768...43H}     & \underline{0.100 29}4 71          & \underline{0.088 87}4 693         & \underline{182}3.183 2       \\
Mars & 2nd order (this paper)         & \underline{0.100} 384             & \underline{0.088} 918             & \underline{182}6.2           \\
Mars & 4th order (this paper)         & \underline{0.100 291} 836         & \underline{0.088 8}69 977         & \underline{1822.8}71         \\
Mars & 6th order (this paper)         & \underline{0.100 291 642} 754     & \underline{0.088 870 80}1 531     & \underline{1822.865 5}33     \\
Mars & Numerical (this paper)         & \underline{0.100 291 642 478 822} & \underline{0.088 870 803 521 489} & \underline{1822.865 525 162} \B \\
\hline
Neptune & \cite{2011PEPI..187..364S}  & \underline{0.2}10 19              & \underline{0.1}51 47              & \underline{6}241.0           \T \\
Neptune & \cite{1978ppi..book.....Z}  & \underline{0.209} 658             & \underline{0.143} 515             & \underline{61}88.92          \\
Neptune & \cite{2013ApJ...768...43H}  & \underline{0.209} 658 98          & \underline{0.143} 515 34          & \underline{61}88.926 7       \\
Neptune & 2nd order (this paper)      & \underline{0.2}11 151             & \underline{0.143} 688             & \underline{6}264.3           \\
Neptune & 4th order (this paper)      & \underline{0.209} 638 699         & \underline{0.143 4}43 899         & \underline{617}7.9           \\
Neptune & 6th order (this paper)      & \underline{0.209 59}9 221 310     & \underline{0.143 453} 888 914     & \underline{6175}.757 821     \\
Neptune & Numerical (this paper)      & \underline{0.209 597 812 191 680} & \underline{0.143 453 902 049 024} & \underline{6175.678 534 586} \B \\
\hline
Uranus 2 & \cite{2011PEPI..187..364S} & \underline{0.21}4 73              & \underline{0.1}41 60              & \underline{5}801.4           \T \\
Uranus 2 & \cite{1978ppi..book.....Z} & \underline{0.213} 648             & \underline{0.115} 655             & \underline{56}80.32          \\
Uranus 2 & \cite{2013ApJ...768...43H} & \underline{0.213} 648 98          & \underline{0.115} 655 64          & \underline{56}80.324 2       \\
Uranus 2 & 2nd order (this paper)     & \underline{0.21}5 683             & \underline{0.115} 731             & \underline{5}773.8           \\
Uranus 2 & 4th order (this paper)     & \underline{0.213} 642 242         & \underline{0.115 59}3 114         & \underline{566}7.7           \\
Uranus 2 & 6th order (this paper)     & \underline{0.213 57}8 684         & \underline{0.115 596 6}90 381     & \underline{5664}.394 033     \\
Uranus 2 & Numerical (this paper)     & \underline{0.213 576 194 544 737} & \underline{0.115 596 602 475 207} & \underline{5664.265 380 457} \B \\
\end{tabular}
\end{center}
\caption{
The shape of the two layers is computed using several methods for the test cases in Table~\ref{tab:two_layer_test_cases}.
The \cite{2011PEPI..187..364S} values are computed using the method by \cite{2010JGRE..11512003K}.
\cite{1978ppi..book.....Z} refers to 3rd order values as computed by \cite{2011PEPI..187..364S} using the theory in \cite{1978ppi..book.....Z}.
Our values are obtained using the recursive formulas with orders 2 to 6 described in \S\ref{SEC:ANALYTICAL} and Appendix~\S\ref{SEC:HIGH_RECURSIVE}, 
and then using the numerical method described in \S\ref{SEC:NUMERICAL}.
Using our numerical values as reference, which should be exact to the last digit (no rounding), we have underlined the digits which agree with it, to roughly assess their accuracy.
}
\label{tab:two_layer_comparison}
\end{table*}

\begin{table*}
\begin{center}
\begin{tabular}{cccl|llll}
$Q_V$ & $\rho_1/\rho_2$ & $\varepsilon_2$ & Method & $e_{p1}$ & $e_{p2}$ & $e_{q1}$ & $e_{q2}$ \T \B \\
\hline
0.1 & 0.5 & 0.001 & 2nd order & \underline{0.110} 686         & \underline{0.097} 654         & \underline{0.095} 857         & \underline{0.084} 570         \T \\
0.1 & 0.5 & 0.001 & 4th order & \underline{0.110 548 7}62     & \underline{0.097 59}0 596     & \underline{0.095 95}2 320     & \underline{0.084 68}2 230     \\
0.1 & 0.5 & 0.001 & Numerical & \underline{0.110 548 771 238} & \underline{0.097 591 141 031} & \underline{0.095 953 221 967} & \underline{0.084 683 153 224} \B \\
\hline
0.2 & 0.3 & 0.01  & 2nd order & \underline{0.27}5 437         & \underline{0.23}3 621         & \underline{0.23}8 535         & \underline{0.20}2 322         \T \\
0.2 & 0.3 & 0.01  & 4th order & \underline{0.272 7}64 535     & \underline{0.232} 579 902     & \underline{0.239 4}44 944     & \underline{0.203} 668 851     \\
0.2 & 0.3 & 0.01  & Numerical & \underline{0.272 712 086 703} & \underline{0.232 608 245 285} & \underline{0.239 498 516 335} & \underline{0.203 734 133 006} \B \\
\end{tabular}
\end{center}
\caption{Similar to Table~\ref{tab:two_layer_comparison} but for 2-layer synchronous moons.
All three methods are from this paper: 
2nd order from \S\ref{SEC:MULTI_MOONS},
4th order from Appendix \S\ref{SEC:HIGH_RECURSIVE},
and numerical from \S\ref{SEC:NUMERICAL}.
}
\label{tab:two_layer_sats}
\end{table*}

Additionally, several 2-layer test cases are used both in \cite{2011PEPI..187..364S} and in \cite{2013ApJ...768...43H},
so we include a Table to provide our estimates along theirs.
The input values used in each test case are listed in Table~\ref{tab:two_layer_test_cases}
and the results in Table~\ref{tab:two_layer_comparison}.
We find that our analytical and numerical approaches are in very good agreement with each other,
with the recursive values approaching the numerical ones as the order of the analytical expressions increases.
We find the values computed using the methods in \cite{1978ppi..book.....Z} and \cite{2013ApJ...768...43H}
show consistently an accuracy better than our 2nd order but worse than our 4th order solutions.
In comparison, the results of \cite{2011PEPI..187..364S} have an inconsistent accuracy:
better than our 2nd order but worse than our 4th order solutions for the Mars case,
which has the larger core and the slowest rotation of the three cases,
but otherwise significantly worse than our 2nd order solutions for the Neptune and Uranus~2 cases.
Overall we consider satisfactory the agreement of our solutions with \cite{1978ppi..book.....Z} and with \cite{2013ApJ...768...43H}
but find the method by \cite{2010JGRE..11512003K} to suffer from significant inaccuracies
in the limit of a small core.
We note that the theory in \cite{2010JGRE..11512003K} explicitly distinguishes between small core and large core,
so it is entirely possible that the small core theory shows issues while leaving the large core theory unaffected.

Finally, in Table~\ref{tab:two_layer_sats} we provide a few 2-layer test cases for synchronous moons,
which can be used as benchmark and future reference,
since we could not find similar cases to compare to in the literature.

\section{Ceres\label{SEC:CERES}}

Ceres is the largest main-belt object, to be visited in 2015 by the Dawn mission \citep{2007EM&P..101...65R}.
Its low density and fast rotation cause a significant polar flattening,
consistent with the shape of an oblate spheroid,
as determined by stellar occultation \citep{1987Icar...72..507M},
HST observations \citep{2005Natur.437..224T},
and Keck AO observations \citep{2008A&A...478..235C}.
The shape determination by \cite{2005Natur.437..224T} in particular
is based on a full longitudinal coverage of Ceres' rotation,
and measured the semi-axes $a=487.3 \pm 1.8$~km and $c=454.7 \pm 1.6$~km.
The mass of Ceres has been measured by \cite{2008CeMDA.100...27B} in the context of
gravitational perturbations by large main-belt asteroids
in astrometric data, obtaining
$(4.75 \pm 0.03) \times 10^{-10}$~$M_\text{Sun}$,
in agreement with previous estimates
\citep{2000A&A...360..363M,Standish2001,2005SoSyR..39..176P,2006Icar..182...23K}.
The bulk density is then $2.09 \pm 0.02$~g/cm$^3$.
The rotation period of $9.074170 \pm 0.000002$~h was measured by \cite{2007Icar..188..451C} 
using lightcurve data covering a period of almost 50 years.

We have performed a forward modeling Monte Carlo,
using 2-layer (core, crust) and 3-layer (core, mantle, crust) 
numerical models and uniformly sampling the density and volume of each layer.
The numerical approach was preferred to avoid the small errors introduced by analytic solutions, see \S\ref{SEC:COMPARISON} and Table~\ref{tab:two_layer_comparison}.
We impose on each solution the hydrostatic equilibrium condition and determine the exterior shape.
The resulting shape semi-axes $a$ and $c$ and total mass of Ceres
of each solution are then compared to the observed values using the $\chi^2$ statistics 
with 3 degrees of freedom and confidence level CL of 0.50 or 0.95,
the former to determine the parameters of the model which most closely match the nominal shape and mass of Ceres,
the latter to find a wider range of parameters which are more broadly compatible with the observations.

\begin{figure}
\begin{center}
\includegraphics*[width=\columnwidth]{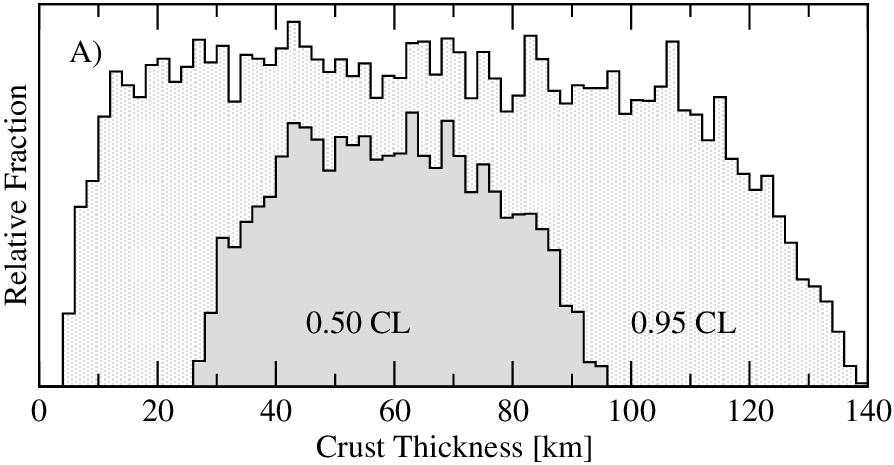}
\includegraphics*[width=\columnwidth]{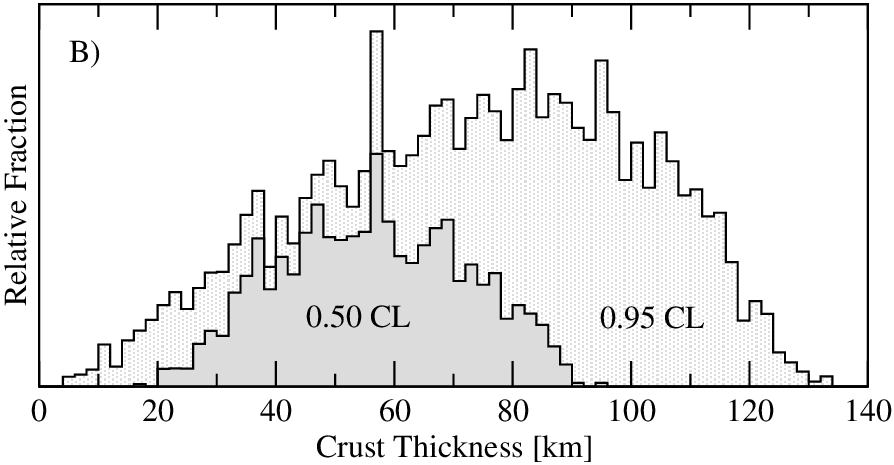}
\end{center}
\caption{Distribution of Ceres 2-layer (A) and 3-layer (B) model solutions versus thickness of the water-ice crust,
at a confidence level (CL) of 0.50 (dark grey) and 0.95 (light grey).
The relative fraction is the ratio of the number of solutions in each bin with the number of baseline
solutions in the same bin, and is in arbitrary linear units.
}
\label{fig:CeresCrustThickness}
\end{figure}

\begin{figure}
\begin{center}
\includegraphics*[width=\columnwidth]{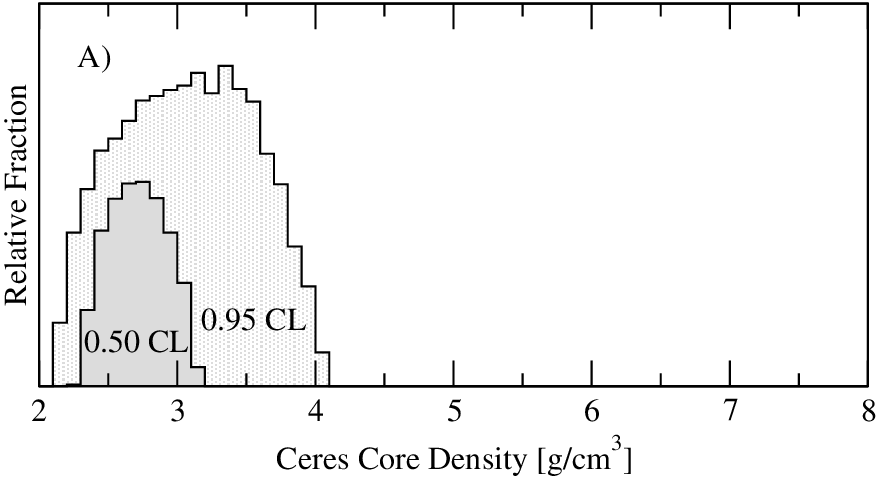}
\includegraphics*[width=\columnwidth]{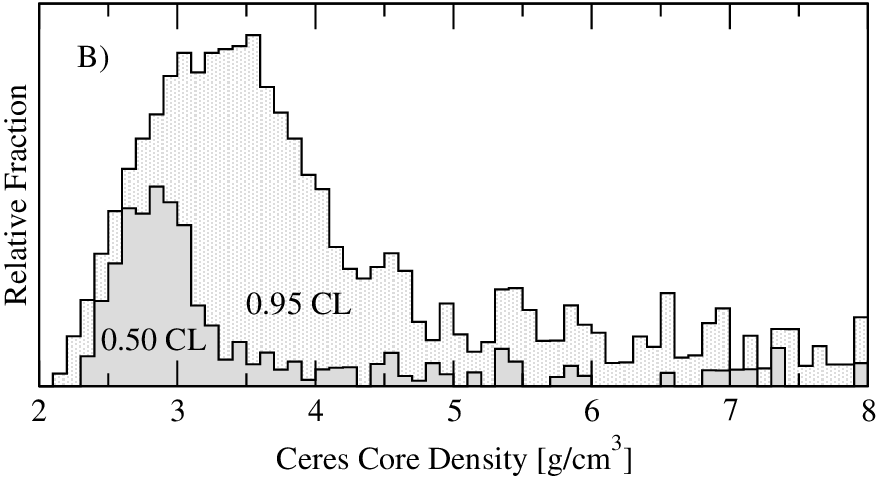}
\end{center}
\caption{Distribution of Ceres 2-layer (A) and 3-layer (B) model solutions versus core density,
same conventions as Figure~\ref{fig:CeresCrustThickness}.
}
\label{fig:CeresCoreDensity}
\end{figure}

Solutions are selected according to our general understanding of the composition of Ceres 
\citep{2005JGRE..110.5009M,2005Natur.437..224T,2009Icar..204..183Z,2010Icar..205..443C}.
The outer crust is assumed to be water-ice, with a density of 0.90--0.95~g/cm$^3$ \citep{Lide_2005},
allowing for a small margin for possible porosity or impurities.
The mantle is assumed to be rocky, with a density of 2.1--3.5~g/cm$^3$.
The core is allowed to have a density of 2.1--8.0~g/cm$^3$,
from a light rocky core up to a possible metallic core.
For Monte Carlo normalization purposes,
we also select a baseline set of solutions 
with global density 0.9--8.0~g/cm$^3$
and 0.99~CL.
For the 2-layer model, a total of 309,814 solutions were included in the baseline set,
with 7,971 included in the 0.95 CL set, 
and 2,885 included in the 0.50 CL set.
For the 3-layer model, a total of 459,662 solutions were included in the baseline set,
with 3,279 included in the 0.95 CL set, 
and 1,223 included in the 0.50 CL set.

\begin{figure}
\begin{center}
\includegraphics*[width=\columnwidth]{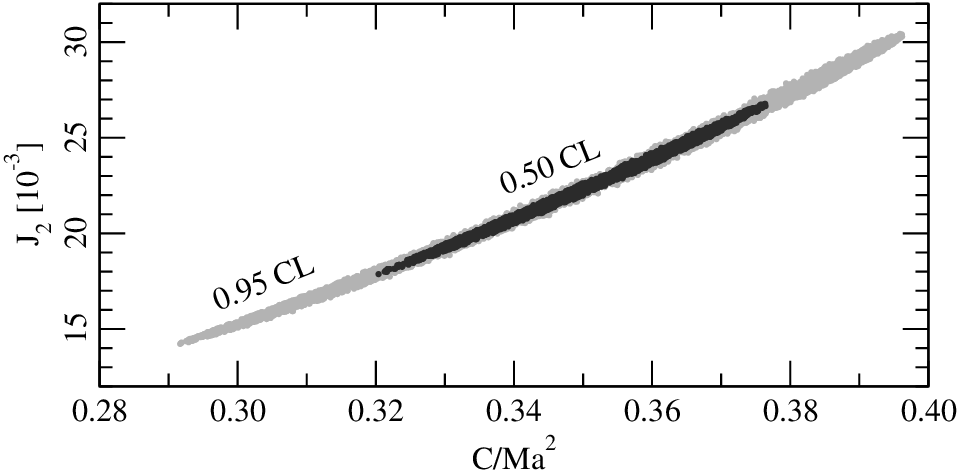}
\end{center}
\caption{Scatter plot of the $J_2$ gravity coefficient versus normalized principal moment of inertia $C/M a^2$
for Ceres solutions, which have virtually identical distributions for the 2-layer and the 3-layer models. 
The 0.50 and 0.95 CL solutions are marked by black and grey regions, respectively.
}
\label{fig:CeresJ2Inertia}
\end{figure}

We find that interior solutions are not compatible with a homogeneous Ceres
at 0.95 CL, in agreement with \cite{2005Natur.437..224T}.
Most of the 2-layer and 3-layer solutions at 0.50 CL show a crust thickness approximately 30--90~km, 
see Fig.~\ref{fig:CeresCrustThickness},
and a core density 2.4--3.1~g/cm$^3$,
see Fig.~\ref{fig:CeresCoreDensity}.
At 0.95 CL the two models show more different solutions:
a crust thickness 
5--130~km for the 2-layer model,
and 20--120~km for the 3-layer model;
a core density 
2.2--4.0~g/cm$^3$ for the 2-layer model,
and 2.4--4.7~g/cm$^3$ for the 3-layer model.
Most 3-layer solutions have a small density difference between core and mantle, typically smaller than 1.0~g/cm$^3$,
which may indicate different levels of hydration \citep{2010Icar..205..443C}.
A low tail of solutions with a metallic core is present in 3-layer solutions at 0.95 CL,
which becomes negligible at 0.50 CL.
In Fig.~\ref{fig:CeresJ2Inertia} we show a scatter plot of 
the resulting $J_2$ gravity coefficients and normalized principal inertia moment $C$.

\section{Solar System Moons\label{SEC:APPMOONS}}

\begin{figure}
\begin{center}
\includegraphics*[width=\columnwidth]{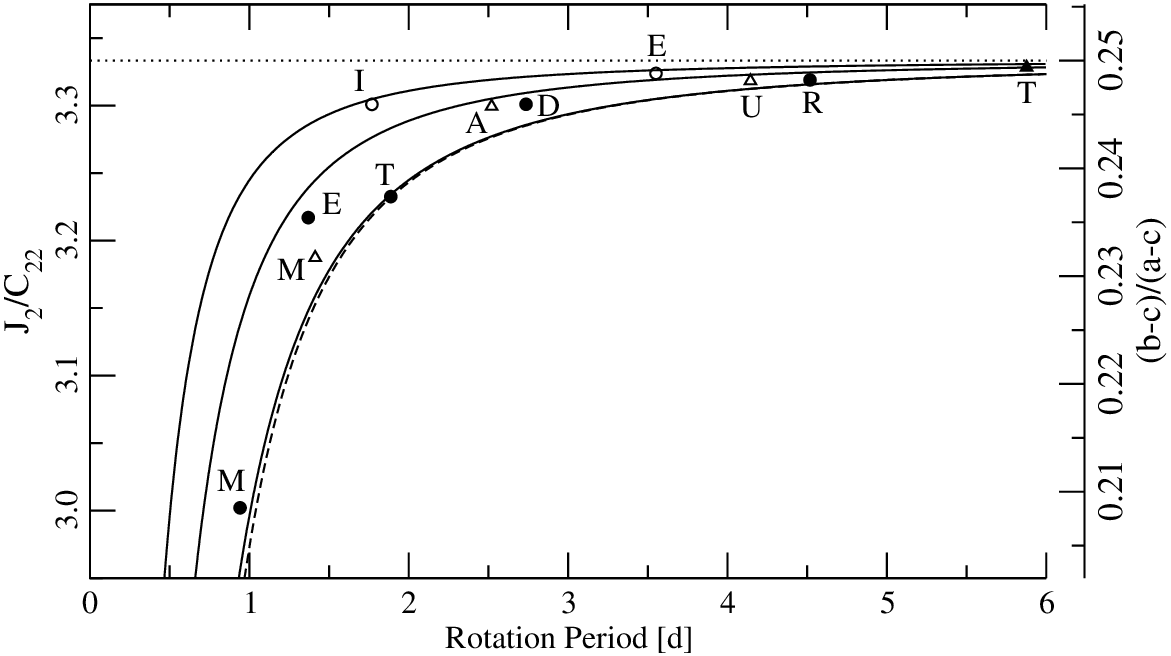}
\end{center}
\caption{Minimum equilibrium values for the $J_2/C_{22}$ and $(b-c)/(a-c)$ ratios versus rotation period.
The solid curves show the numerical solution for homogeneous synchronous moons of density 1, 2, and 4~g/cm$^3$.
The dashed curve shows the approximated solution from Eq.~\eqref{eq:bcac_exp2} and \eqref{eq:J2_C22_exp2} for density 1~g/cm$^3$.
Moons of Jupiter$^a$ (open circle): Io, Europa.
Moons of Saturn$^b$ (solid circle): Mimas, Enceladus, Tethys, Dione, Rhea.
Moons of Uranus$^c$ (open triangle): Miranda, Ariel, Umbriel.
Moons of Neptune$^d$ (solid triangle): Triton.
References for bulk density data:
$^a$\cite{2004jpsm.book..281S};
$^b$\cite{2010Icar..208..395T};
$^c$\cite{1992AJ....103.2068J};
$^d$\cite{2000Icar..148..587T}.
}
\label{fig:tidal_ratios}
\end{figure}

This method can be applied to solar system moons, and 
in Figure~\ref{fig:tidal_ratios} we have plotted the minimum value of
the $J_2/C_{22}$ and $(b-c)/(a-c)$ ratios 
for selected bulk density values, and included the estimated value 
for several large and fast-rotating
moons of the giant planets of the solar system.
These values are computed numerically using the observed rotation period and bulk density.
For reference, the asymptotic values of $J_2/C_{22}=10/3$ and $(b-c)/(a-c)=1/4$
are given by the dotted line.

The decrease of the ratio $(b-c)/(a-c)$ for fast rotators was already present in 
\cite{1969efe..book.....C} and noted in \cite{1988Icar...73...25D}.
The analytical dependence to order $\Lambda^2$ of the ratios 
$J_2/C_{22}$ and $(b-c)/(a-c)$
can be obtained using the relations in Eq.~\eqref{eq:tidal_high_first} to \eqref{eq:tidal_high_last},
and in the limit of an homogeneous synchronous moon ($\sigma_2 \to 0$)
we obtain the compact expressions:
\begin{align}
e_{p1}^2 & = \frac{15}{2} \Lambda^2 - \frac{1125}{112} \Lambda^4 + \cdots \label{eq:ep1sq_homogeneous_lambda} \\
e_{q1}^2 & = \frac{45}{8} \Lambda^2 + \frac{3375}{448} \Lambda^4 + \cdots \label{eq:eq1sq_homogeneous_lambda}
\end{align}
and the ratio
\begin{align}
\frac{b-c}{a-c} & = \frac{1}{4} - \frac{1485}{896} \Lambda^2 + \cdots
\label{eq:bcac_exp2}
\end{align}
while for the gravity coefficients we have
\begin{align}
J_2                & = \frac{15}{16} \Lambda^2 - \frac{2475}{896} \Lambda^4 + \cdots \\
C_{22}             & = \frac{9}{32} \Lambda^2 + \frac{675}{1792} \Lambda^4 + \cdots
\end{align}
with the ratio
\begin{align}
\frac{J_2}{C_{22}} & = \frac{10}{3} - \frac{100}{7} \Lambda^2 + \cdots
\label{eq:J2_C22_exp2}
\end{align}
These expressions for the ratios to order $\Lambda^2$
are good approximations for all solar system moons,
with the largest relative error of about 1\% for Mimas and much smaller for all the other moons,
see Figure~\ref{fig:tidal_ratios}.

\section{Discussion}
\label{SEC:DISCUSSION}

We suggest a Ceres icy crust thickness of 30--90~km,
which is in good agreement with thermal evolution scenarios by \cite{2005JGRE..110.5009M}
and \cite{2010Icar..205..443C}.
\cite{2005Natur.437..224T} determine an icy crust thickness of 66--124~km using a 2-layer model,
which tends to be larger than the values suggested here. 
This may be due to the modeling of the shape of the interior layers,
and it is not clear how the method by \cite{1993Icar..105..326T} 
which is used in \cite{2005Natur.437..224T} deals with this issue.
Our core density estimate of 2.4--3.1~g/cm$^3$ agrees with the 2.7~g/cm$^3$ by \cite{2005Natur.437..224T}.

The $J_2/C_{22}$ ratio in synchronous moons is often constrained a-priori to its nominal value of $10/3$
when determining the individual gravity coefficients from flyby radio science data
\citep{2003PEPI..136..201A,2004jpsm.book..281S,2007GeoRL..34.2202A,2008GeoRL..35.5204M,2010PEPI..178..176A}.
As we show in Eq.~\eqref{eq:J2_C22_exp2} and Fig.~\ref{fig:tidal_ratios},
the value of $J_2/C_{22}$ has a significant correction due to fast rotation,
and this should be taken into account when determining the gravity coefficients.
A similar correction is present in the $(b-c)/(a-c)$ ratio,
see Eq.~\eqref{eq:bcac_exp2} and Fig.~\ref{fig:tidal_ratios}.

The shape of planets and synchronous moons is expected to 
viscously relax and asymptotically reach the equilibrium shape,
which may or may not be reached within the age of the solar system
depending on how effective the relaxation process is
\citep{1973Icar...18..612J}.
Of six well-measured Saturnian large icy moons, three (Tethys, Dione, Rhea) have clearly interpretable equilibrium shapes, 
and the other three (Mimas, Enceladus, Iapetus) appear to be significantly off hydrostatic equilibrium 
\citep{2007Icar..190..573T,2010Icar..208..395T},
and this can be due to their formation mechanisms, 
as well as their thermal, dynamical, rotational, and collisional history.
The analytical and numerical solutions presented
can help us assess how far moons are from hydrostatic equilibrium,
potentially providing an insight on the different scenarios leading to their current shapes.

\section{Conclusions}

The analytical and numerical solutions presented in this manuscript
mark a clear improvement over previous methods.
The 2nd order recursive analytical relations 
in Eq.~\eqref{eq:rot_N_epsq} for planets 
and in Eq.~\eqref{eq:tidal_N_epsq} and \eqref{eq:tidal_N_eqsq} for synchronous moons apply to an arbitrary number of layers,
and express in a very compact form how the shape of each layer is a generalized weighted average of the shapes of all the other layers,
with the relative sizes and densities as weights,
plus a source term proportional to $\Omega^2$.
High-order relations can be obtained for bodies with a small number of layers,
as we did for 2-layer bodies in Appendix \S\ref{SEC:HIGH_RECURSIVE}.
Finally, the numerical method allows to obtain solutions which are exact
within the precision of the floating point operations,
and converge for slow- and fast-rotating bodies,
up to a number of layers which is essentially limited by
the processing power available.

Our results have important applications to solar system planets and synchronous moons.
For planets, accounting for the deformation of each layer is a significant improvement over previous models,
and when applied to Ceres, 
it generates solutions which have a water-ice crust thinner than previously thought.
For synchronous moons, 
we model the deformation due to rotation and tidal effects jointly, 
extending the range of applicability of analytical solutions to moderately fast rotators,
and obtaining 2nd order analytical expressions for the $J_2/C_{22}$ and $(b-c)/(a-c)$ ratios.

\acknowledgments

This research was supported by the NASA DAVPS program, grant NNX10AR20G.
We thank Mark Sykes and two anonymous referees for providing useful comments.

\appendix

\section{Gravitational Potential of a Homogeneous Triaxial Ellipsoid}

The gravitational potential of an homogeneous triaxial ellipsoid (TE) is analytic \citep{book_macmillan}:
\begin{align}
\begin{split}
\frac{U_\text{TE}(x,y,z)}{\pi G \rho}
& = \frac{2 a b c}{\sqrt{a^2-c^2}} \left[ 1 - \frac{x^2}{a^2-b^2} + \frac{y^2}{a^2-b^2} \right] F(\psi,e_q^2/e_p^2) \\
& + \frac{2 a b c}{\sqrt{a^2-c^2}} \left[ \frac{x^2}{a^2-b^2} - \frac{(a^2-c^2) y^2}{(a^2-b^2)(b^2-c^2)} + \frac{z^2}{b^2-c^2} \right] E(\psi,e_q^2/e_p^2) \\
& + \frac{2 a b c}{\sqrt{(a^2+\kappa)(b^2+\kappa)(c^2+\kappa)}} \left[ \frac{c^2+\kappa}{b^2-c^2} y^2 - \frac{b^2+\kappa}{b^2-c^2} z^2 \right]
\label{eq:U_TE}
\end{split}
\end{align}
where
$\sin^2(\psi) = (a^2-c^2)/(a^2+\kappa)$,
$F$ and $E$ are the elliptic integral functions of the first and second kind, respectively.
The variable $\kappa$ is the positive root of the equation
\begin{align}
\frac{x^2}{a^2+\kappa} + \frac{y^2}{b^2+\kappa} + \frac{z^2}{c^2 + \kappa} = 1
\label{eq:kappa}
\end{align}
for points $(x,y,z)$ outside the ellipsoid,
and $\kappa=0$ inside.
The expansion for $e_p \ll 1$ is:
\begin{align}
\begin{split}
\frac{U_\text{TE,in}(x,y,z)}{\pi G \rho} 
& = \frac{2}{3}(3a^2-r^2) - \frac{2}{15}(5a^2-r^2+3z^2) e_p^2 - \frac{2}{15}(5a^2-r^2+3y^2) e_q^2 \\
& - \frac{2}{105}(14a^2-4r^2+12z^2) e_p^4 + \frac{2}{105}(14a^2+r^2-3x^2) e_p^2 e_q^2 - \frac{2}{105}(14a^2-4r^2+12y^2) e_q^4 + \cdots
\label{eq:U_TE_in_exp}
\end{split} \\
\begin{split}
\frac{U_\text{TE,out}(x,y,z)}{\pi G \rho} 
& = \frac{4 a^3}{3 r} - \frac{2 a^3 \left[5r^4-a^2(r^2-3z^2)\right]}{15 r^5} e_p^2 - \frac{2 a^3 \left[5r^4-a^2(r^2-3y^2)\right]}{15 r^5} e_q^2 \\
& - \frac{a^3 \left[ 35r^8 +14 a^2 r^4 (r^2-3 z^2)-3 a^4 (3 x^4+6 x^2 (y^2-4 z^2)+3 y^4-24 y^2 z^2+8 z^4) \right]}{210 r^9} e_p^4 \\
& + \frac{a^3 \left[ 35r^8+7 a^2 r^4 (-2 x^2+y^2+z^2)  + 3 a^4 (x^4-3 x^2 (y^2+z^2)-4 y^4+27 y^2 z^2-4 z^4)\right]}{105 r^9} e_p^2 e_q^2 \\
& - \frac{a^3 \left[ 35r^8 +14 a^2 r^4 (r^2-3 y^2)-3 a^4 (3 x^4+6 x^2 (z^2-4 y^2)+3 z^4-24 y^2 z^2+8 y^4) \right]}{210 r^9} e_q^4 + \cdots 
\label{eq:U_TE_out_exp}
\end{split}
\end{align}
where $r^2=x^2+y^2+z^2$ and the potential is continuous with continuous derivative at the surface of the triaxial ellipsoid.
For $\kappa$ we have:
\begin{align} 
\kappa & = r^2 - a^2 + \frac{a^2 z^2}{r^2} e_p^2 + \frac{a^2 y^2}{r^2} e_q^2 + \frac{a^4 (x^2+y^2) z^2}{r^6} e_p^4 - 2 \frac{a^4 y^2 z^2}{r^6} e_p^2 e_q^2 + \frac{a^4 (x^2+z^2) y^2}{r^6} e_q^4 +  \cdots
\label{eq:kappa_sol_TE}
\end{align}
and a recursive formula can be obtained simply by extracting a $\kappa$ factor from Eq.~\eqref{eq:kappa}:
\begin{align} 
\kappa & \longleftarrow \frac{x^2}{1+a^2/\kappa} + \frac{y^2}{1+b^2/\kappa} + \frac{z^2}{1+c^2/\kappa}
\end{align}
and using $\kappa = r^2 - a^2$ as initial value.

\section{Gravitational Potential of a Homogeneous Oblate Spheroid}

The gravitational potential takes a simpler form in the case of oblate spheroids (OS), 
where $e_q=0$ \citep{book_macmillan}:
\begin{align}
\begin{split}
\frac{U_\text{OS}(x,y,z)}{\pi G \rho} 
& = \frac{2 a^2 c}{\sqrt{a^2-c^2}} \left[ 1 - \frac{r^2-3z^2}{2(a^2-c^2)} \right] \psi + \frac{a^2 c}{a^2-c^2} \left[ \frac{\sqrt{c^2+\kappa}}{a^2+\kappa} (x^2+y^2) - \frac{2z^2}{\sqrt{c^2+\kappa}} \right]
\label{eq:U_OS}
\end{split}
\end{align}
and the expansion of $U_\text{OS}$ for $e_p \ll 1$ can be recovered directly 
from the $U_\text{TE}$ expressions in Eq.~\eqref{eq:U_TE_in_exp} and \eqref{eq:U_TE_out_exp}
by simply setting $e_q=0$.

\section{Spherical Harmonics \label{SEC:SH}}

The gravitational potential of an homogeneous triaxial ellipsoid
can be expanded in spherical harmonics \citep{1966tsga.book.....K,1995geph.conf....1Y} to obtain:
\begin{align}
U = \frac{G M}{r} \left[ 1 + \sum_{l=2}^{\infty} \sum_{m=0}^{l} \frac{a_1^l}{r^l} C_{lm} P_{lm}(\cos(\theta)) \cos(m\phi) \right] \label{eq:potential}
\end{align}
where
$P_{lm}(\cos(\theta))$ is the associate Legendre function,
$\cos(\theta)=z/r$ and $\tan\phi=y/x$.
This series converges absolutely outside the sphere with reference radius $a_1$,
and the unnormalized coefficients $C_{lm}$ 
can be determined by integrating 
over the volume of the body \citep{book_macmillan,1995geph.conf....1Y}.
The coefficients are \citep{book_macmillan,1995geph.conf....1Y,2008CeMDA.100..319T}:
\begin{align}
\begin{split}
\label{eq:Clmi}
C_{lm} 
& =
\frac{(2-\delta_{0m})}{2^l}
\frac{(l-m)!}{(l+m)!}
\sum_{p=0}^{l/2} 
\sum_{q=0}^{m/2} 
(-1)^{p+q}
\binom{l}{p} 
\binom{2l-2p}{l} 
\binom{m}{2q} 
\\
&
(l - m - 2 p + 1)_{m}
\sum_{\nu_x=0}^p
\sum_{\nu_y=0}^{p-\nu_x}
\frac{p!}{\nu_x!\nu_y!(p-\nu_x-\nu_y)!}
{\cal N}_{m-2q+2\nu_x,2q+2\nu_y,l-m-2\nu_x-2\nu_y}
\end{split}
\end{align}
where both $l$ and $m$ are even, and $C_{lm}=0$ otherwise.
The generalized moments of inertia coefficients ${\cal N}_{n_x n_y n_z}$ are: 
\begin{align}
{\cal N}_{n_x n_y n_z} 
& = 
\frac{3}{4\pi} 
\dfrac{\Gamma\left[(n_x+1)/2\right]\Gamma\left[(n_y+1)/2\right]\Gamma\left[(n_z+1)/2\right]}{\Gamma\left[(n_x+n_y+n_z+5)/2\right]}
(1-e_q^2)^{n_y/2}
(1-e_p^2)^{n_z/2}
\end{align}
if $n_x,n_y,n_z$ are all even, and ${\cal N}_{n_x n_y n_z}=0$ otherwise.
The non-zero $C_{lm}$ terms for $l \leq 4$ are:
\begin{align}
C_{20} & = \frac{(e_q^2-2e_p^2)}{10} &
C_{22} & = \frac{e_q^2}{20} &
C_{40} & = \frac{3(8 e_p^4 - 8 e_p^2 e_q^2 + 3 e_q^4)}{280} &
C_{42} & = \frac{(e_q^4 - 2 e_p^2 e_q^2)}{280} &
C_{44} & = \frac{e_q^4}{2240}
\end{align}
and in the special case of an oblate spheroid ($e_q=0$) we have that only the 
$C_{l0}$ terms with even $l$ are non-zero, with
\begin{align}
C_{l0} & = (-1)^{l/2} \frac{3}{(l+1)(l+3)} e_p^l
\end{align}
and using the convention $J_l=-C_{l0}$ we have
\begin{align}
J_2 & = \frac{1}{5} e_p^2 &
J_4 & = - \frac{3}{35} e_p^4 &
J_6 & = \frac{1}{21} e_p^6 &
J_8 & = - \frac{1}{33} e_p^8 &
J_{10} & = \frac{3}{143} e_p^{10}
\end{align}
When computing the $C_{lm}$ for a multi-layer body,
we have to include the relative size factor $a_i^l/a_1^l$ 
to refer all contributions to the same exterior semi-axis $a_1$,
and also scale by the mass fraction $M_i / M = V_i (\rho_i-\rho_{i-1}) / M$
contributed by the layer, 
with $V_i = (4\pi/3) a_i^3 (1-e_{pi}^2)^{1/2} (1-e_{qi}^2)^{1/2}$ volume of the layer, 
and $M = \sum_{i=1}^{N} M_i = \sum_{i=1}^{N} V_i (\rho_i-\rho_{i-1})$ total mass of the body,
to obtain
$C_{lm} = \sum_{i=1}^{N} (a_i/a_1)^l (M_i/M) C_{lm,i}$
where $C_{lm,i}$ is the single layer $C_{lm}$ from Eq.~\eqref{eq:Clmi}.

\section{Inertia Moments \label{SEC:INERTIA}}

The principal moments of inertia $A,B,C$ for a homogeneous ellipsoid are:
\begin{align}
\frac{A}{M a_1^2} & = \frac{2-e_{p1}^2-e_{q1}^2}{5} &
\frac{B}{M a_1^2} & = \frac{2-e_{p1}^2}{5} &
\frac{C}{M a_1^2} & = \frac{2-e_{q1}^2}{5} 
\end{align}
and when combining it in a multi-layer body, we have:
\begin{align}
\frac{A}{M a_1^2} & = \sum_{i=1}^{N} \frac{a_i^2}{a_1^2} \frac{M_i}{M} \frac{(2-e_{pi}^2-e_{qi}^2)}{5} &
\frac{B}{M a_1^2} & = \sum_{i=1}^{N} \frac{a_i^2}{a_1^2} \frac{M_i}{M} \frac{(2-e_{pi}^2)}{5} &
\frac{C}{M a_1^2} & = \sum_{i=1}^{N} \frac{a_i^2}{a_1^2} \frac{M_i}{M} \frac{(2-e_{qi}^2)}{5} 
\end{align}

\section{Pressure}

The interior pressure $p$ of a multi-layer body in hydrostatic equilibrium 
can be obtained from Eq.~\eqref{eq:hydro}:
\begin{align}
p(x,y,z) = \sum_{i=1}^{N(x,y,z)} (\rho_i-\rho_{i-1}) \left[ U_\text{tot}(x,y,z) - U_\text{tot}(a_i,0,0) \right]
\end{align}
where the sum is over the $N(x,y,z)$ layers containing the point $(x,y,z)$.

\section{Inverse Maclaurin Relation \label{SEC:INVML}}

The angular velocity $\Omega$ and polar eccentricity $e_p$ of a rotating homogeneous fluid body in hydrostatic equilibrium follow the exact relation \citep{1969efe..book.....C}
\begin{align}
\frac{\Omega^2}{\pi G \rho} & = \Lambda^2 = \frac{(1-e_p^2)^{1/2}}{e_p^3} 2 (3-2e_p^2) \arcsin(e_p) - \frac{6}{e_p^2}(1-e_p^2)
\end{align}
which can be expanded in power series for $e_p \ll 1$ as
\begin{align}
\Lambda^2 & = \frac{8}{15}e_p^2 + \frac{8}{105}e_p^4 - \frac{64}{3465}e_p^8 - \frac{1024}{45045}e_p^{10} + \cdots
\label{eq:MLser}
\end{align}
and this relation can be inverted using a recursive scheme 
\begin{align}
e_p^2 & \longleftarrow \frac{15}{8} \left[ \Lambda^2 - \frac{8}{105}e_p^4 + \frac{64}{3465}e_p^8 + \frac{1024}{45045}e_p^{10} + \cdots \right]
\label{eq:inv_base_iter}
\end{align}
where the value of $e_p^2$ on the left side of Eq.~\eqref{eq:inv_base_iter} is successively improved 
by substituting the value of the expression on the right side.
Starting with $e_p^2=0$,
we get $e_p^2 = (15/8) \Lambda^2$ after the first iteration,
and after five iterations we have:
\begin{align}
e_p^2 & = \frac{15}{8}\Lambda^2 - \frac{225}{448}\Lambda^4 + \frac{3375}{12544}\Lambda^6 + \frac{3830625}{15454208}\Lambda^8 + \frac{349565625}{803618816}\Lambda^{10} + \cdots 
\label{eq:invMLser}
\end{align}

\section{High-Order Recursive Solutions \label{SEC:HIGH_RECURSIVE}}

The equations in recursive form including 6th order eccentricity terms 
for a 2-layer planet are
\begin{align}
\begin{split}
e_{p1}^2 & \longleftarrow \frac{15\Lambda^2 + 12 \mu_2^5 \sigma_2 e_{p2}^2}{8 + 20 \mu_2^3 \sigma_2}
           - \frac{ (8 + 105 \mu_2^3 \sigma_2) e_{p1}^4 - (70 \mu_2^3 + 84 \mu_2^5) \sigma_2 e_{p1}^2 e_{p2}^2 + (42 \mu_2^5 + 15 \mu_2^7) \sigma_2 e_{p2}^4}{7(8 + 20 \mu_2^3 \sigma_2)} \\
         & \phantom{\longleftarrow} - \frac{175 \mu_2^3 \sigma_2 e_{p1}^6  - (105 \mu_2^3 + 210 \mu_2^5) \sigma_2 e_{p1}^4 e_{p2}^2  - (35 \mu_2^3 - 84 \mu_2^5 - 120 \mu_2^7) \sigma_2 e_{p1}^2 e_{p2}^4  + (21 \mu_2^5 - 15 \mu_2^7 - 35 \mu_2^9 )\sigma_2 e_{p2}^6  }{14(8 + 20 \mu_2^3 \sigma_2)} + \cdots
\end{split} \\
\begin{split}
e_{p2}^2 & \longleftarrow \frac{15\Lambda^2 + 12 e_{p1}^2}{20 + 8\sigma_2} 
           + \frac{ 48 e_{p1}^4 - 56 e_{p1}^2 e_{p2}^2 - 8 \sigma_2 e_{p2}^4}{7(20 + 8 \sigma_2)} 
           + \frac{32 e_{p1}^6}{7(20 + 8 \sigma_2)} + \cdots
\end{split}
\end{align}

For a 2-layer synchronous moon we have the 4th order expansion:
\begin{align}
\begin{split}
e_{p1}^2 & \longleftarrow \frac{60\Lambda^2 + 12 \mu_2^5 \sigma_2 e_{p2}^2}{8 + 20 \mu_2^3 \sigma_2} 
           - \frac{(22 + 140 \mu_2^3 \sigma_2) e_{p1}^4 - (70 \mu_2^3 + 105 \mu_2^5) \sigma_2 e_{p1}^2 e_{p2}^2 + (42 \mu_2^5 + 15 \mu_2^7 ) \sigma_2 e_{p2}^4}{7(8 + 20 \mu_2^3 \sigma_2)} \label{eq:tidal_high_first} \\
         & \phantom{\longleftarrow} + \frac{16 e_{p1}^2 e_{q1}^2 + (70 \mu_2^3 - 42 \mu_2^5) \sigma_2 e_{p1}^2 e_{q2}^2 - (42 \mu_2^5 - 30 \mu_2^7) \sigma_2 e_{p2}^2 e_{q2}^2}{7(8 + 20 \mu_2^3 \sigma_2)} + \cdots
\end{split} \\
e_{p2}^2 & \longleftarrow \frac{60\Lambda^2 + 12 e_{p1}^2}{20 + 8\sigma_2} + \frac{48 e_{p1}^4 - 35 e_{p1}^2 e_{p2}^2 - (35 + 22 \sigma_2) e_{p2}^4}{7(20 + 8\sigma_2)} 
           - \frac{12 e_{p1}^2 e_{q1}^2 - 28 e_{q1}^2 e_{p2}^2 - 16 \sigma_2 e_{p2}^2 e_{q2}^2}{7(20 + 8 \sigma_2)} + \cdots \\
\begin{split}
e_{q1}^2 & \longleftarrow \frac{45\Lambda^2 + 12 \mu_2^5 \sigma_2 e_{q2}^2}{8 + 20 \mu_2^3 \sigma_2} 
           - \frac{(8 + 105 \mu_2^3 \sigma_2) e_{q1}^4 + (42 \mu_2^5 + 15 \mu_2^7) \sigma_2 e_{q2}^4}{7(8 + 20 \mu_2^3 \sigma_2)} \\
         & \phantom{\longleftarrow} + \frac{16 e_{p1}^2 e_{q1}^2 + (70 \mu_2^3 - 42 \mu_2^5) \sigma_2 e_{q1}^2 e_{p2}^2 + (70 \mu_2^3 + 84 \mu_2^5) \sigma_2 e_{q1}^2 e_{q2}^2 - (42 \mu_2^5 - 30 \mu_2^7) \sigma_2 e_{p2}^2 e_{q2}^2}{7(8 + 20 \mu_2^3 \sigma_2)} + \cdots
\end{split} \\
e_{q2}^2 & \longleftarrow \frac{45\Lambda^2 + 12 e_{q1}^2}{20 + 8\sigma_2} + \frac{48 e_{q1}^4 - 8 \sigma_2 e_{q2}^4}{7(20 + 8\sigma_2)} - \frac{12 e_{p1}^2 e_{q1}^2 + 56 e_{q1}^2 e_{q2}^2 - 28 e_{p1}^2 e_{q2}^2 - 16 \sigma_2 e_{p2}^2 e_{q2}^2}{7(20 + 8\sigma_2)} + \cdots \label{eq:tidal_high_last}
\end{align}

\section{Equipotential Surface Outside an Homogeneous Triaxial Ellipsoid}
\label{SEC:DEVIATION}

An homogeneous triaxial ellipsoid with semi-major axis $a_1$ 
and polar and equatorial eccentricities $e_{p1}$ and $e_{q1}$
generates an equipotential surface at its exterior,
which can be approximated by triaxial ellipsoid with semi-major axis $a_0>a_1$
and eccentricities $e_{p0}$ and $e_{q0}$.
By using Eq.~\eqref{eq:U_TE_out_exp}
and imposing $U_\text{TE,out}(a_0,0,0) = U_\text{TE,out}(0,b_0,0) = U_\text{TE,out}(0,0,c_0)$
we can obtain the 4th order recursive expressions
\begin{align}
e_{p0}^2 & \longleftarrow \frac{3}{5} \frac{a_1^2}{a_0^2} e_{p1}^2 + \frac{177}{700} \frac{a_1^4}{a_0^4} e_{p1}^4 + \frac{6}{175} \frac{a_1^4}{a_0^4} e_{p1}^2 e_{q1}^2 - \frac{3}{4} e_{p0}^4 + \cdots \\
e_{q0}^2 & \longleftarrow \frac{3}{5} \frac{a_1^2}{a_0^2} e_{q1}^2 + \frac{177}{700} \frac{a_1^4}{a_0^4} e_{q1}^4 + \frac{6}{175} \frac{a_1^4}{a_0^4} e_{p1}^2 e_{q1}^2 - \frac{3}{4} e_{q0}^4 + \cdots
\end{align}
which do not depend on the density or angular velocity of the homogeneous ellipsoid, and converge to
\begin{align}
e_{p0}^2 & = \frac{3}{5} \frac{a_1^2}{a_0^2} e_{p1}^2 - \frac{3}{175} \frac{a_1^4}{a_0^4} e_{p1}^4 + \frac{6}{175} \frac{a_1^4}{a_0^4} e_{p1}^2 e_{q1}^2 + \cdots \\
e_{q0}^2 & = \frac{3}{5} \frac{a_1^2}{a_0^2} e_{q1}^2 - \frac{3}{175} \frac{a_1^4}{a_0^4} e_{q1}^4 + \frac{6}{175} \frac{a_1^4}{a_0^4} e_{p1}^2 e_{q1}^2 + \cdots
\end{align}

The equipotential surface is close to but not exactly an ellipsoid,
and using spherical coordinates 
$(x,y,z) = \left( r \sin(\theta)\cos(\phi), r \sin(\theta)\sin(\phi), r \cos(\theta) \right)$
we can calculate the ratio $\eta = {r_U}/{r_E}$,
where $r_U$ is the radius of the equipotential surface which is obtained by imposing $U_\text{TE,out}(x,y,z) = U_\text{TE,out}(a_0,0,0)$,
while $r_E$ is the radius of the ellipsoid
\begin{align}
\frac{1}{r_E^2} & = \frac{\sin^2(\theta) \cos^2(\phi)}{a_0^2} + \frac{\sin^2(\theta)\sin^2(\phi)}{a_0^2(1-e_{q0}^2)}  + \frac{\cos^2(\theta)}{a_0^2(1-e_{p0}^2)}
\end{align}
The expression for $\eta$ is then
\begin{align}
\eta & = 1 - \frac{3}{200} \frac{a_1^4}{a_0^4} \left[ \sin^2(2\theta) e_{p1}^4 - 2 \sin^2(2\theta) \sin^2(\phi) e_{p1}^2 e_{q1}^2 + \left( 3+\cos(2\theta)+2\cos(2\phi)\sin^2(\theta) \right) \sin^2(\theta) \sin^2(\phi) e_{q1}^4  \right] + \cdots
\label{eq:deviation_from_ellipsoid}
\end{align}
showing that the deviation of the equipotential surface from the reference ellipsoid
includes only terms of 4th order in eccentricity or higher,
and decreases as the 4th power of the distance.
Also, we have that $\eta \le 1$ which means that the equipotential surface
is in general just interior to the reference ellipsoid.
In particular, $\eta=1$ at the extremes of the principal axes for a triaxial ellipsoid,
while for an oblate spheroid $\eta=1$ at the poles and at the equator.

The minimum of $\eta$ is in the limit of $a_0 \to a_1$
and shows a complex angular dependence. 
For a homogeneous rotating planet in hydrostatic equilibrium,
with $e_{p1}$ from Eq.~\eqref{eq:invMLser} and $e_{q1}=0$,
we have that $\eta$ from Eq.~\eqref{eq:deviation_from_ellipsoid} becomes
\begin{align}
\eta & = 1 - \frac{27}{512} \frac{a_1^4}{a_0^4} \sin^2(2\theta) \Lambda^4 + \cdots
\end{align}
with equal minima at $\theta=\pi/4$ and at $\theta=3 \pi/4$,
which correspond to latitudes of $\pm 45^\circ$.
For an homogeneous synchronous moon in hydrostatic equilibrium,
with  
$e_{p1}$ from Eq.~\eqref{eq:ep1sq_homogeneous_lambda} 
and 
$e_{q1}$ from Eq.~\eqref{eq:eq1sq_homogeneous_lambda},
we have that $\eta$ from Eq.~\eqref{eq:deviation_from_ellipsoid} becomes
\begin{align}
\eta & = 1 - \frac{27}{2048} \frac{a_1^4}{a_0^4} \left[ (77 + 59 \cos(2\theta)) \sin^2(\theta) - 18\sin^4(\theta)\cos(4\phi) + 30\sin^2(2\theta)\cos(2\phi) \right] \Lambda^4 + \cdots
\end{align}
with equal minima along the 4 specularly symmetric directions
$(\theta,\phi) = \left\{ (\pi/4,0), (\pi/4,\pi), (3\pi/4,0), (3\pi/4,\pi) \right\}$
where
\begin{align}
\eta & = 1 - \frac{27}{32} \frac{a_1^4}{a_0^4} \Lambda^4 + \cdots
\end{align}

\end{document}